# Exploring the effect of the spatial scale of fishery management

Nao Takashina[1,*] and Marissa L. Baskett[2]

1. *Department of Biology, Faculty of Sciences, Kyushu University, 6-10-1, Hakozaki, Fukuoka, 8128581, Japan*
   *nao.takashina@gmail.com*
2. *Department of Environmental Science and Policy, University of California, Davis, One Shields Ave, Davis, CA 95616*
   *mlbaskett@ucdavis.edu*

## Abstract

For any spatially explicit management, determining the appropriate spatial scale of management decisions is critical to success at achieving a given management goal. Specifically, managers must decide how much to subdivide a given managed region: from implementing a uniform approach across the region to considering a unique approach in each of one hundred patches and everything in between. Spatially explicit approaches, such as the implementation of marine spatial planning and marine reserves, are increasingly used in fishery management. Using a spatially explicit bioeconomic model, we quantify how the management scale affects optimal fishery profit, biomass, fishery effort, and the fraction of habitat in marine reserves. We find that, if habitats are randomly distributed, the fishery profit increases almost linearly with the number of segments. However, if habitats are positively autocorrelated, then the fishery profit increases with diminishing returns. Therefore, the true optimum in management scale given cost to subdivision depends on the habitat distribution pattern.

**Keywords**: bioeconomic model, management scale, and marine reserves

## Introduction

The importance of spatial scale has been well recognized in many fields of ecology (Levin 1992), such as species-area relationships, maps of species richness, and conservation planning (Palmer and White 1994, Schwartz 1999, Margules and Pressey 2000, Turner and Tjørve 2005, Hurlbert and Jetz 2007). Spatially explicit approaches to ecosystem management introduce a management scale overlaid on the natural spatial scale of ecological processes. Specifically, managers must decide how much to subdivide the area under concern: from implementing a uniform approach across the region to considering a unique approach in each of hundreds of patches and everything in between.

This scale of management assessment and implementation affects the ability to achieve management goals. For example, analysis of range-map data at inappropriately fine resolutions might lead to the identification of erroneous "biodiversity hotspots" with overly optimistic estimates of species representation in reserves and potentially invalid complementarity sets for identifying conservation priorities (Hurlbert and Jetz 2007).

Ecosystem-based fisheries management (EBFM) is an inherently spatially explicit approach to fisheries management, including the implementation of marine reserves, or no-take zones (Pikitch et al. 2004). Marine reserves' goals range from conserving species to support sustainable fisheries management (Leslie 2005, Lester et al. 2009). Even without reserves, EBFM typically involves a spatially explicit approach to harvest decision in terms of zonal allocations of fishing effort (Francis et al. 2007), which can increase fishery profit over spatially uniform management if appropriately based on habitat distribution and connectivity (Rassweiler et al. 2012). However, few studies explicitly considered the effect of the choice of spatial scale in spatial fishery management on achieving management goals.

Under spatial fisheries management, managers must choose a management scale to define a management unit (i.e., zoning unit), and fishing regulations such as entry limitation and establishment of reserves occur within these zoning units (Cancino et al. 2007, White and Costello 2011). For example, the concept of setting variable harvest rates over space was implemented for co-occurring fisheries of less productive and productive species in the US west coast, such as yellowtail and canary rockfish (Francis 1986) and yelloweye rockfish and lingcod (Dougherty et al. 2013). Spatial management through a fine filter enables managers to allocate fishing efforts and reserves more flexibly compared to management through a coarse filter, but a finer filter imposes greater complexity on the decision-making process and enforcement. For territorial user rights fisheries (TURFs), coarser management scales increase achievement of optimal harvest due to the greater degree of ownership and lower competition (White and Costello 2011). However, for fisheries under top-down control such as the case where federal-level government decisions determine individual fishing effort, the appropriate management scale might change because competition between management units does not occur.

To investigate how the choice of spatial management scale affects fishery and ecological outcomes such as optimal fishery profit, biomass, fishery effort, and the fraction of habitat in marine reserves, we construct a spatially explicit bioeconomic model that follows an age-structured harvested population. Using two California species, cabezon (*Scorpaenichthys marmorata*) and red abalone (*Haliotis rufescens*), we compare two spatial management strategies: allocating reserve or non-reserve patches with a uniform fishing rate versus allocating fishing rate in each patch, where allocation within the management scale maximizes fishery profit. We then investigate the relationship between the spatial scale of management and the above-mentioned fishery and ecological outcomes under varying degrees of autocorrelation in the habitat, which determines the spatial scale of habitat.

## Methods

We aim to construct the simplest possible model that allows us to quantify the relationship between the choice of spatial management scale and our metrics for fishery and ecological outcomes. As detailed below, the managed population occurs in a naturally patchy habitat, where the choice of management scale relative to the natural habitat patch size determines its effect on population dynamics. We explore different values of spatial autocorrelation in habitat patches to model different levels of natural patchiness. Larval dispersal connects the patches, where populations then experience density-dependent recruitment. Post-settlement individuals remain within habitat patches (i.e., a relatively sedentary species) according to an age-structured model with density-independent natural and harvest mortality; the structured population dynamics allow us to determine the effect of management decisions on population biomass and biomass yield. To model top-down control given a particular management scale, the fishery optimizes profit across the entire habitat based on management-patch-specific effort allocation, with two approaches. First, management patches have either zero effort (reserves) or harvest, with the same effort in all harvested patches and both this effort level and which patches are harvested are chosen to maximize yield (uniform effort, or UE, strategy); this approach models the optimal use of reserves in fishery management, with no further spatially-explicit management beyond reserve designation. Second, the amount of effort in each management patch (including the possibility of zero effort) is chosen to maximize yield (fine-tuned effort, or FE, strategy); this approach models a fully spatially explicit management approach. We then determine the effect of management scale on effort and profit as our fishery outcomes as well as population biomass and fraction of the habitat in marine reserves as our ecological outcomes.

### Environmental and management scale

The target species population occurs along a coastline where we approximate the geographic landscape by a one-dimensional patchy environment with different patterns of autocorrelation in habitat quality. The minimum size of habitat defines the environmental scale that determines the population dynamics. Whether or not fishing occurs in a given location depends on a separate management scale (Fig. 1). We define the management scale as the size of a minimum management unit where fishing effort is uniform within the region. We assume that the minimum management scale is the environmental scale. The environmental scale inherently depends on ecological and physiological characteristics of a species and geomorphological patterns (Levin 1992). The management scale depends on managers or fishermen based on, for example, assessment data or range maps (Hopkinson et al. 2000, Hurlbert and Jetz 2007), and it characterizes the spatial fishery management. Here we set the environmental scale (minimum habitat patch size) to 1km to match the minimum environmental scale of the target species in their post-larval home range (i.e. <1km adult movement). The management scale can then be $2^0$, $2^1$, $\cdots$, or $2^n$ times larger than the environmental scale where $n$ is the number of subdivisions. Hereafter, we use "habitat patch", or, briefly, "patch" to indicate the environmental scale, and we explicitly refer to "management patch" when discussing a management location.

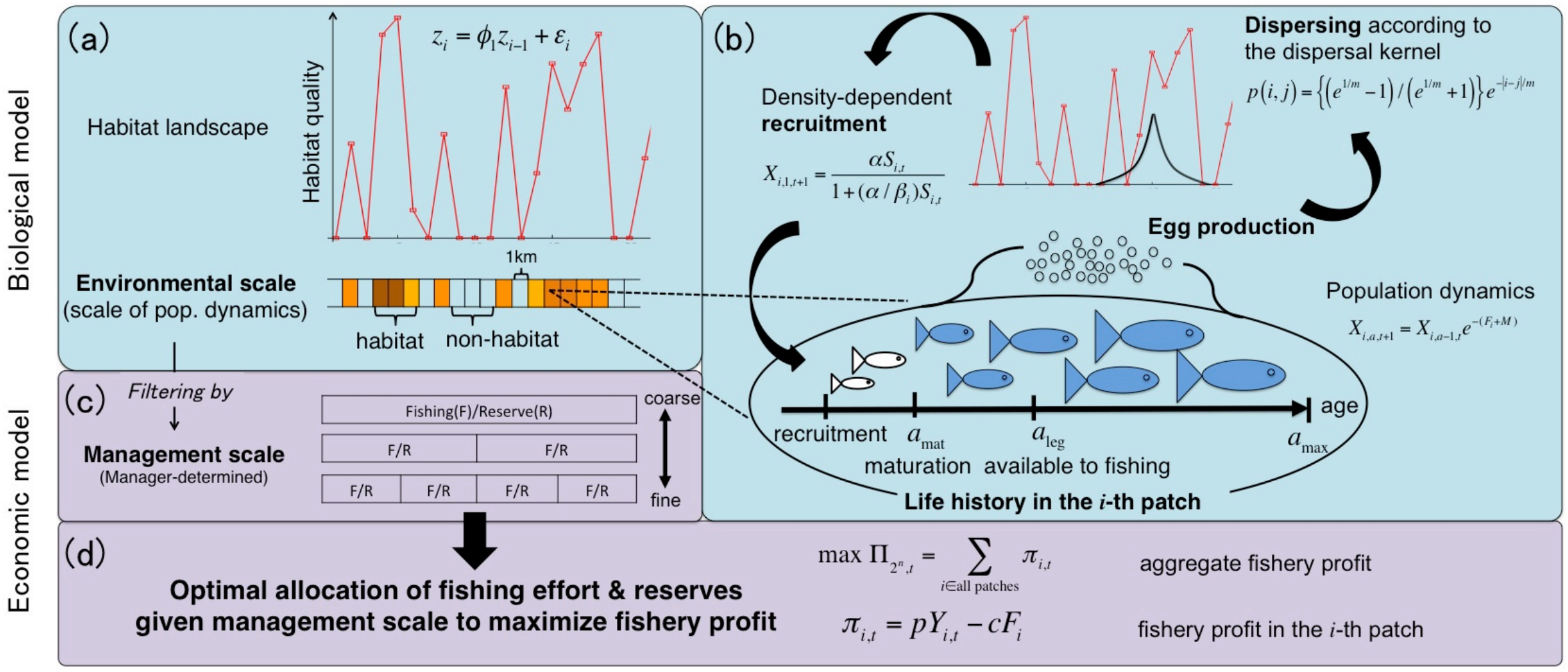

**Figure 1** Schematic description of the model. (a) The degree of autocorrelation determines the environmental scale. (b) Population dynamics occur at the environmental scale. Planktonic larval dispersal connects individual patches. Larvae successfully arriving at a patch experience density-dependent recruitment and subsequently follow age-structured dynamics. (c) Managers chose a management scale for a given region. (d) Managers allocate fishing efforts and reserves to each management patch so as to maximize fishery profit for each given management scale.

## Habitat landscape

We construct an $n$-patch habitat landscape $\left(\text{relative carrying capacity } z_1, \cdots, z_n\right)$ with the auto-regressive model (AR(1) model) to generate various autocorrelations patterns between patches in a manner analogous to how they are often measured in field data (Dale and Fortin 2009). Increasing positive autocorrelation indicates increasing similarity between neighboring patches and therefore increasing habitat scale. With the degree of autocorrelation $\phi_1$ and white noise $\varepsilon_i$ with 0 mean and a variance of 1, we employ the AR(1) model

$$z_i = \phi_1 z_{i-1} + \varepsilon_i . \tag{1}$$

We regard the *i*-th patch to be habitat if $z_i > 0$ and non-habitat if $z_i \le 0$. Specifically we define $z_i$ as $z_i = \max\left(\phi_1 z_{i-1} + \varepsilon_i,\ 0\right)$, and any individuals that disperse to non-habitat do not survive.

**Population dynamics**

In the bioeconomic model underlying our analysis, post-settlement age classes experience natural mortality at a rate *M* and, after growing to the age at legal size $a_{\text{leg}}$, fishing mortality at a patch-specific rate $F_i$. The dynamics for the population abundance $X_{i,a,t}$ in the *i*-th patch at age *a* and time *t* (year) in each time step are then

$$X_{i,a,t+1} = \begin{cases} X_{i,a-1,t} e^{-M}, & 2 \le a < a_{\text{leg}} \\ X_{i,a-1,t} e^{-(F_i+M)}, & a_{\text{leg}} \le a \le a_{\max}, \end{cases} \tag{2}$$

given maximum age $a_{\max}$.

We convert age to size to both calculate population biomass as one of our output metrics and to calculate larval production in the population dynamics. We obtain the length at age $L_a$ using the von Bertalanffy growth equation, given the asymptotic length $L_\infty$, the age at 0 cm $a_0$, and growth rate *k*: $L_a = L_\infty\left(1 - e^{-k(a-a_0)}\right)$. We then obtain the biomass for each age $W_a$ using the allometlic relationship with constants $b_1$ and $b_2$, $W_a = b_1 L_a^{b_2}$.

Total biomass in the system at time $t$, $B_t$, is the sum of the biomass over all age classes and patches, $B_t = \sum_i \sum_a W_{i,a} X_{i,a,t}$. For reproduction, we convert age to size to fecundity $E_a$, where sexually matured individuals (age $a \geq a_{\text{mat}}$) produce larvae after the fishing season. In cabezon, we use a fecundity-at-weight relationship to calculate $E_a$ (Table 1), and in abalone we use a fecundity-at-length relationship to calculate $E_a$ (Table A1). Total reproductive output is then $R_{i,t} = \sum_{a\text{mat}}^{a\max} E_a X_{i,a,t}$.

We assume a sessile post-settlement stage where all connectivity between patches occurs through larval dispersal. To model this dispersal we introduce a discretized analogue of the Laplacian kernel, which represents the probability $p(i,j)$ that a larva is transported from $i$-th patch to $j$-th patch by ocean currents. Specifically, we start with an exponentially decreasing probability with distance from the larval origin $p(i,j) \propto e^{-|i-j|/m}$ given integers $i$ and $j$ (Botsford et al., 2001; Kaplan, 2006; Lockwood et al., 2002), and choose the proportionality coefficient so as to satisfy $\sum_{j=-\infty}^{j=\infty} p(i,j) = 1$. The settlement probability is then:

$$p(i,j) = \left\{ \left(e^{1/m} - 1\right) / \left(e^{1/m} + 1\right) \right\} e^{-|i-j|/m}, \tag{3}$$

where $m$ is a non-dimensional quantity that determines the migration ability of the species, such that the mean dispersal distance is $\bar{m} = 2e^{1/m} / \left(e^{1/m} + 1\right)\left(e^{1/m} - 1\right)$. All larvae that settle outside a habitat patch die, i.e., $p(i,k) = 0$, for $k$ outside the habitat.

Density-dependent Beverton-Holt survivorship occurs among the total number of larvae (settlers) $S_{i,t} = \sum_j R_{j,t} p(i,j)$ arriving at the $i$-th patch at the end of the fishing season in time $t$. Specifically, given the maximum settler survival rate $\alpha$ and the carrying capacity in the $i$-th patch $\beta_i = z_i K$ (where $K$ is the baseline carrying capacity of one environmental patch with the value $10^4$), the number of recruits (age class 1) at time $t$+1 and location $i$ is:

$$X_{i,1,t+1} = \frac{\alpha S_{i,t}}{1+(\alpha/\beta_i)S_{i,t}} . \quad (4)$$

**Table 1** Cabezon (*Scorpaenichthys marmorata*) parameters, following the parameterization in White et al. (2010).

| Parameter | Description | Value | Source* |
|---|---|---|---|
| $L_\infty$ | Maximum size | 62.12 cm | a |
| $k$ | Growth rate | 0.18 cm/year | a |
| $a_0$ | Age at 0 cm | -1.06 year | a |
| $b_1$ | Coefficient in length-to-weight relationship | $9.2\times10^{-6}$ | a |
| $b_2$ | Exponent in length-to-weight relationship | 3.187 | a |
| $E_a$ | Fecundity-at-weight | $(15.3W_a+27.3)\times10^3$ eggs | b |
| $a_{\max}$ | Maximum age | 15 years | a, c |
| $a_{\text{mat}}$ | Age at maturity | 3 years | a |
| $a_{\text{catch}}$ | Age available to fishing | 4 years (38.1cm) | d |
| $M$ | Natural mortality rate | 0.25/year | a |
| $\bar{m}$ | Mean larvae dispersal distance | 100 km | c |

* a. Cope & Punt (2005); b. O'Connell (1953); c. White et al. (2010); d. CDFW (2014b)

## Fishery dynamics

The manager's allocation of fishing effort among the management patches depends on the single management scale and the goal of maximizing the equilibrium aggregate fishery profit $\Pi^*_{2^n}$ under the given fishing strategy (see below for two strategies), where $2^n$ indicates the number of management patches. The aggregate fishery profit at time *t*, $\Pi_{2^n,t}$ is the sum of the management patch-specific profits, $\pi_{i,t}$ over all patches:

$$\Pi_{2^n,t} = \sum_i \pi_{i,t} . \qquad (5)$$

The patch-specific fishery profit $\pi_{i,t}$ is the fishery revenue (product of the price *P* and biomass yield in the *i*-th patch at time *t*, $Y_{i,t}$) minus the harvest cost (product of per-patch cost of fishing mortality *c* and fishing effort in the *i*-th patch $F_i$):

$$\pi_{i,t} = PY_{i,t} - cF_i . \qquad (6)$$

Given the biomass of legal-sized fish ($a \geq a_{leg}$) in the *i*-th patch at time *t*, $B_{i,t}^{leg} = \sum_{a\mathrm{leg}}^{a\max} W_a X_{i,a,t} F_i$ , the yield in the *i*-th patch at time *t*, $Y_{i,t}$ , is

$$Y_{i,t} = \frac{B_{i,t}^{leg} F_i \left(1 - e^{-(F_i+M)}\right)}{F_i + M} \qquad (7)$$

(Rassweiler et al., 2012).

We explore two optimal fishing strategies: (i) uniform effort strategy (UE strategy), and (ii) fine-tuned effort strategy (FE strategy). For the UE strategy, optimization occurs via allocation of fishing grounds ($A_i$=1 in management patch *i*) and zero effort or no-take reserve if the patch is habitat ($A_i$=0 to indicate a management patch with zero fishing effort) in each of the $2^n$ management patches given a single fishing effort *F* on all open patches. Then the UE strategy achieves an optimization of aggregate fishery profit by choosing $2^n + 1$ parameters $\left(A_1,\, A_2, \cdots,\, A_{2^n},\, F\right)$ together so as to maximize the objective function at equilibrium:

$$\max_{A,F} \Pi_{2^n}^* = \max_{A,F} \sum \pi_i^* , \qquad (8)$$

where $\pi_i^*$ represents management patch-specific profit at equilibrium. For the FE strategy, optimization occurs via designating a different fishing mortality $F_i$ in each of $2^n$ management patch $\left(F_1,\, F_2, \cdots,\, F_{2^n}\right)$, where $F_i$ =0 in management patches that include habitat patch(es) represent cases where reserves are part of the management strategy. The objective function for the FE strategy is then

$$\max_{F} \sum \pi_i^* . \qquad (9)$$

Because it allows finer tuning of effort control, the FE strategy will inevitably outperform the UE strategy in terms of fishery profit, but it might be more costly to implement (see Discussion). Appendix S1 describes the two algorithms in more detail.

Note that if fishing occurs in a non-habitat patch or unproductive habitat with a small carrying capacity, then it may result in a cost of $cF_i$, or a negative profit in the patch on the environmental scale, but on the management scale all management patches satisfy $\pi_{i,t} \geq 0$. While our investigation of management scale does not account for effort elimination within unprofitable environmental patches as a finer scale management does or as might occur through fisher behavior (e.g., fishers eventually noticing and avoiding non-habitat patches as they fish within a management unit), this is an effect that the manager tends to avoid non-productive habitats given a fixed management scale. Also, the negative profit may serve, in practice, as an approximated cost associated with coarse-filtered management (e.g., ineffective initial searching due to an unspecified/broad management unit scale).

**Model parameterization**

We parameterize the model based on two example species in the California coast, cabezon (*Scorpaenichthys marmorata*) and red abalone (*Haliotis rufescens*), that have a relatively sedentary adult phase (adult home range <1km) and differ in their larval dispersal distance. These two species are economically important nearshore species and are expected to benefit from spatial fisheries management (CDFW 2014a). For simplicity we define *P* (price/kg) as 1 and *c* (cost/fishing effort) as a constant value (3000 in the main text; we investigate the parameter dependence in Appendix S2); our focus is on relative qualitative trends across varying management scales. We present results for cabezon in main text and for red abalone in Appendix S2 for demonstrating qualitatively robust results across these life histories (see Table 1 for cabezon and Appendix Table S1 for red abalone for the parameter values used in the analysis).

**Analysis**

We investigate the effect of the different management scales on the aggregate fishery profit, total population biomass, fishing mortality rate for the UE strategy and average fishing mortality rate over non-reserve patches for the FE strategy, and the fraction of marine reserves, defined as [the number of habitat patches with no fishing mortality]/[the total number of habitat patches]. We consider a coastline containing $2^7$ patches for the environmental scale. Therefore, the feasible set of management scales is $\left\{2^7\text{km}, 2^6\text{km},\cdots, 2^0\text{km}\right\}$ and the corresponding set of the number of management patches is $\left\{1, 2,\cdots, 128\right\}$. We show the average value over 100 simulation trials, where the AR(1) model randomly generates a different landscape in each simulation run. In the simulations, we only consider the practically relevant cases where the fishery is profitable.

## Results

### Spatial fisheries management in positively autocorrelated landscapes

Because we do not account for a cost to subdivision, applying a finer management scale always increases the net benefit of fishing regardless of the fishing strategy (Fig. 2; note that we discuss the reason for this and other outcomes in the Discussion below). When the habitat landscape is positively autocorrelated ($\phi_1 = 0.9$ in Eq. 1), the relative aggregate fishery profit increases with a finer management scale but with diminishing returns (Fig. 2a). The FE strategy always outperforms the UE strategy, but both show the same qualitative trend of diminishing returns with finer management scale. In addition, the effort distribution of the FE strategy is, by definition, more complex than that of the UE strategy (e.g., filed squares and solid line in Fig. 2e vs. 2d).

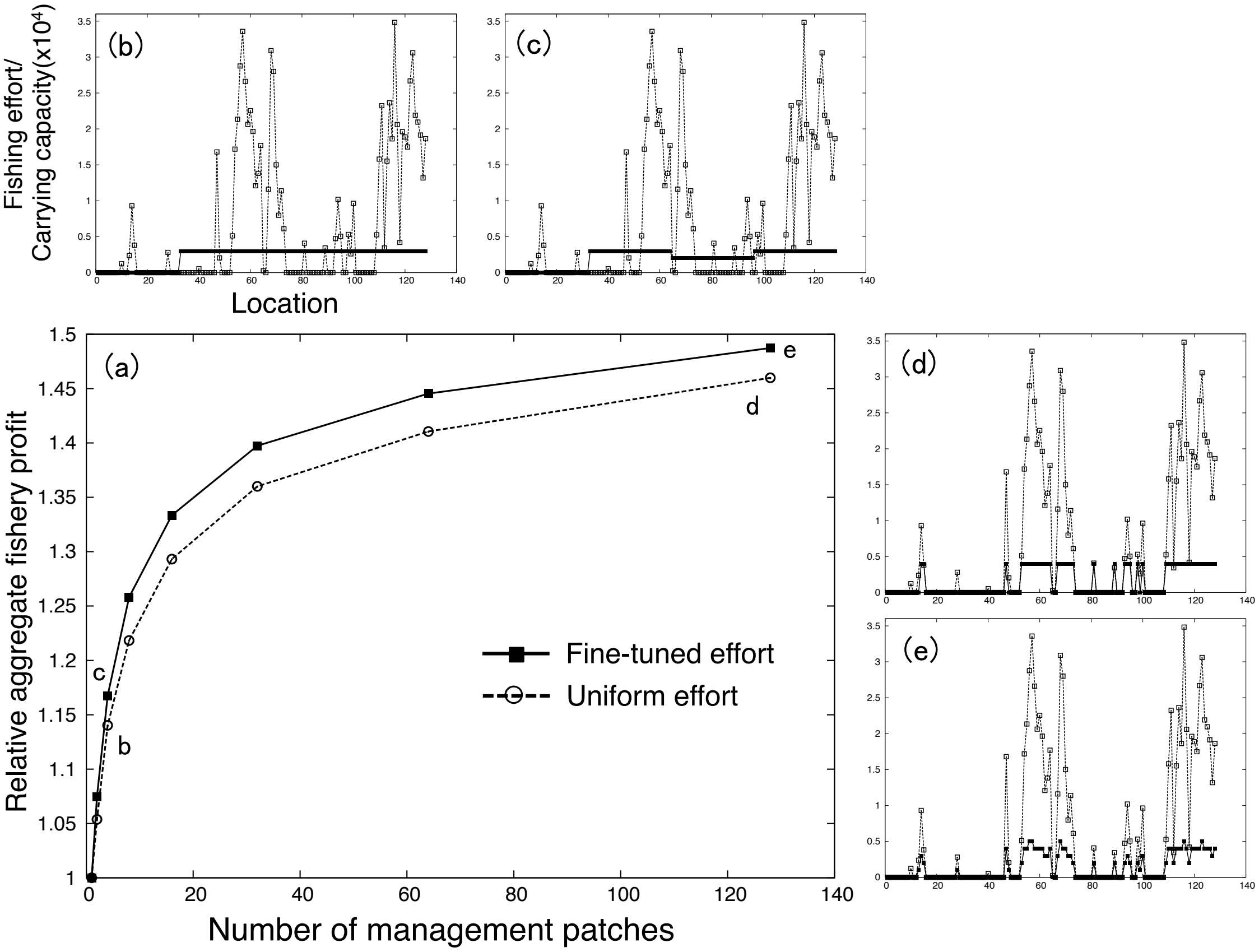

**Figure 2** The effect of management scale on profit in an autocorrelated landscape ($\phi_1 = 0.9$). (a) Aggregate fishery profit, plotted relative to the profit given one management patch, as a function of management scale. Each line represents the fine-tuned effort strategy (FE strategy; squares, solid line) or the uniformed effort strategy (UE strategy; circles, dashed line). (b)-(e) Examples of fine-tuned effort (filed squares, solid line) for sample realizations of habitat landscape (squares, dashed line). Management scales and fishing strategies in each panel correspond to notations in the panel (a).

The total population biomass decreases as management scale becomes finer (Fig. 3a), and the decline is more rapid under the FE strategy than the UE strategy. Initially, the reserve fraction increases rapidly, then it quickly saturates (Fig. 3b). The UE strategy always requires a larger reserve fraction than the FE strategy for optimal profit. Fishing mortality (Fig. 3c) increases with an increasing number of management patches at a decreasing rate. The UE strategy always requires higher fishing mortality rate than the FE strategy for optimizing profit.

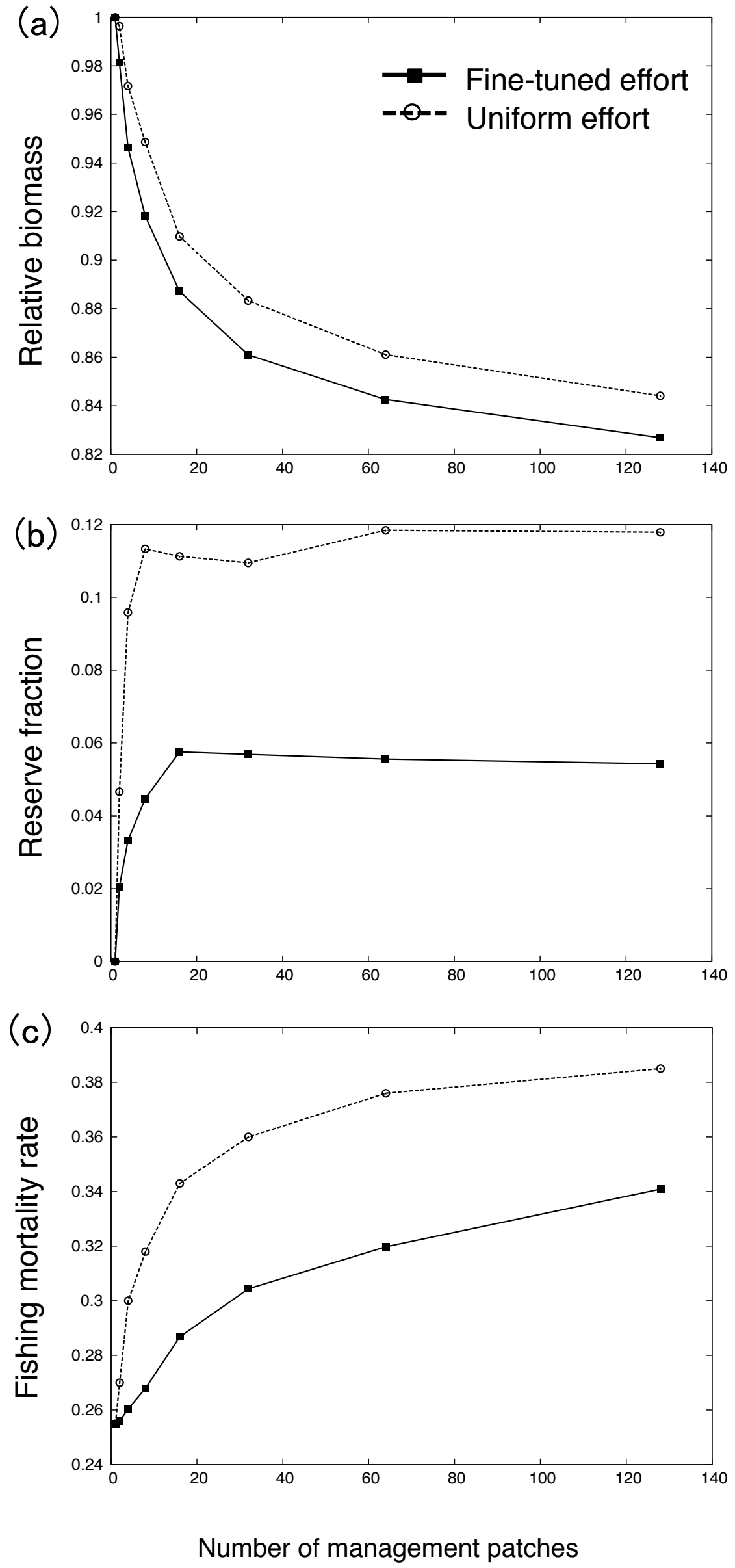

**Figure 3** The effect of management scale on biomass and management controls in an autocorrelated landscape ( $\phi_1 = 0.9$ ) with the FE strategy (squares, solid line) or the UE strategy (circles, dashed line). Each panel shows (a) biomass relative to its value given one management patch, (b) reserve fraction, defined as [the number of reserves]/[the number of habitats], and (c) fishing mortality rate for the UE strategy and average fishing mortality rate over non-reserve patches for the FE strategy. These results are the average value of the 100 times simulation.

**Spatial fisheries management in uncorrelated landscapes**

Without autocorrelation ($\phi_1 = 0$ in Eq. 1; white noise), fishery profit shifts to a near-linear function of management scale (Fig. 4a, FE strategy; the UE strategy shows an analogous qualitative trend). In other words, increasing the number of management patches no longer exhibits diminishing returns in fishery profit. Relative biomass initially increases with an increasing number of management patches when that number is small, but after that it declines almost linearly as the number of management patches increases (Fig. 4b). Reserve fraction also peaks at an intermediate management scale, but at a greater number of patches than biomass (Fig. 4c). Fishing mortality initially decreases with an increasing number of management patches when that number is small, then subsequently increases almost linearly (Fig. 4d).

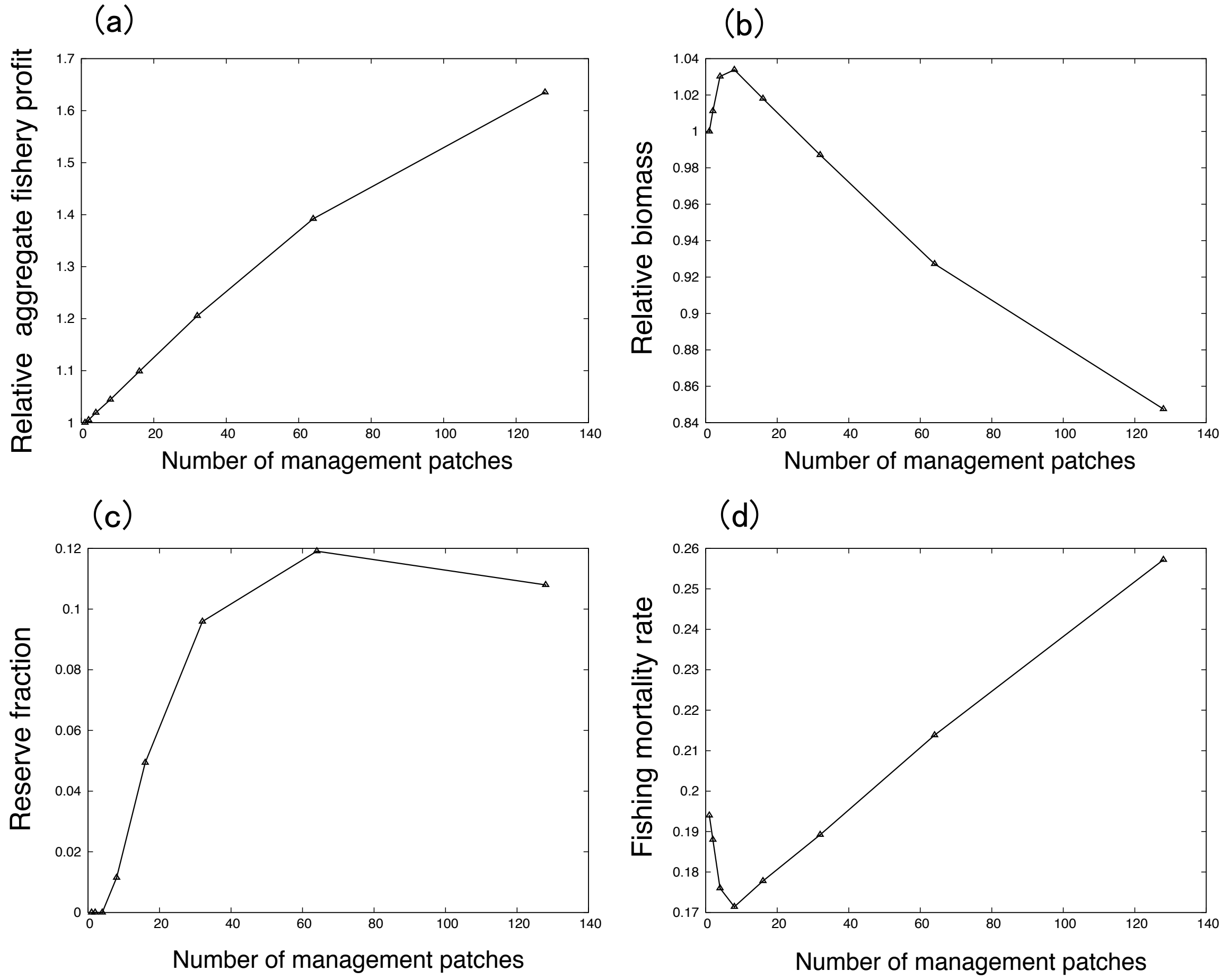

**Figure 4** The effect of management scale in an uncorrelated landscape ( $\phi_1 = 0.0$ ) under the FE strategy. Each panel shows (a) aggregate fishery profit, plotted relative to its value given one management patch, (b) biomass, plotted relative to its value given one management patch, (c) reserve fraction, and (d) average fishing mortality rate. These results are the average value of the 100 times simulation.

**Management efficacy and parameter sensitivity**

Looking beyond aggregate fishery profit to a question of how fisheries can effectively distribute fishing effort, the relative fraction of unprofitable patches declines with an increasing number of management patches (Fig. 5, FE strategy). This decline is more rapid for an autocorrelated landscape than an uncorrelated landscape, where the decline is nearly linear.

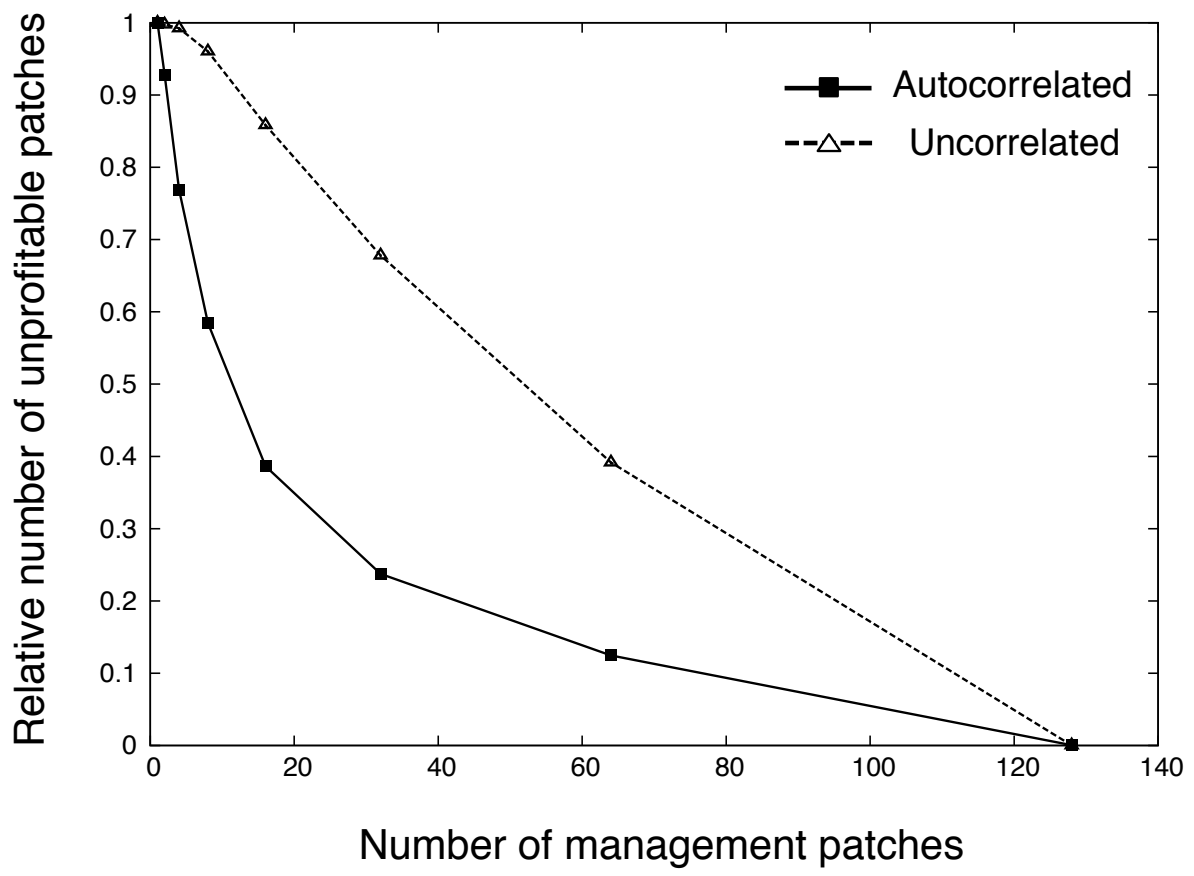

**Figure 5** The number of unprofitable patches in the optimal effort allocation under the FE strategy. They are plotted relative to the number given one management patch, as a function of the environmental scale both in an autocorrelated landscape ( $\phi_1 = 0.9$ , box, thick line) and an uncorrelated landscape ( $\phi_1 = 0.0$ , triangle, dashed line).

The qualitative trends described above are consistent for different values of settler survival, fishing effort cost, and natural mortality (Fig. S1) as well as for red abalone parameter values (Figs. S2, S3).

We assumed the total patch number is 128 and show the average value of the 100 simulation trials, and we verified that 100 trials were sufficient to capture the qualitative trends that drive our conclusions (Appendix S5; Fig. S5a). Note that, however, some effects of the stochasticity in habitat landscape generation remain, which arise from randomly-drawn habitat landscapes with either a large number or small number of habitat patches. Each trial shows quantitatively different results in the values of our output metrics, while the relative values across different management scales typically hold in each trial. The variability in the number of habitat patches of the landscape particularly affects the reserve fraction because the number of reserves primarily depend on the number of habitat patches in the landscape, which causes the within-trend fluctuations in Fig. 3b (see Fig. S5b).

## Discussion

Assuming no cost to management subdivision, fishery profit increases with increasing management patches due to increased flexibility in spatial management. However, the shape of this relationship (which determines the potential for an intermediate peak in profit with management scale given cost to subdivision; discussed below) depends critically on the degree of autocorrelation in the natural habitat. Specifically, profit saturates rapidly with the number of subdivisions with autocorrelation but increases in a near linear relationship without autocorrelation (Figs 2a, 4a). A landscape with positive autocorrelation is more clustered and hence contains larger-sized habitat patches. On the other hand, in uncorrelated landscapes individual habitat patches tend to be dispersed more evenly and averaged habitat size is smaller. Therefore, a finer filter is needed for spatial management in uncorrelated landscapes than autocorrelated landscape to achieve an effective allocation of fishing effort and marine reserves. In other words, if the same management scale is applied, managers inevitably place the fishing effort in a larger amount of unprofitable patches with uncorrelated landscapes than with autocorrelated landscapes (fig. 5). Given these contrasting relationships, the incremental improvement with increasing the number of subdivisions exhibits diminishing returns for autocorrelated habitats (fig. 2a) but is consistent across values for management scale in uncorrelated habitats (fig. 4a).

Given the same degree of autocorrelation in landscapes, fisheries outcomes show qualitatively similar results between two different fishing strategies, the optimized effort allocation (FE) strategy and the uniform effort allocation (UE) strategy (Figs. 2 and 3) which indicates robustness of our results to different fishing approaches. However, in the FE strategy results in greater profit with lower fishing effort and lower population biomass and reserve fraction compared to the UE strategy: the FE strategy allows fishermen to increase fishery profit more effectively through the fine-tuned effort allocation and consequently it causes a larger decline in the biomass in comparison with the UE strategy. Regardless of the optimal zonal allocation of fishing effort, marine reserves are typically a part of optimal management, corresponding to previous models (Neubert 2003, Sanchirico et al. 2006), because the variability in habitat quality, in combination with dispersal, leads to patches where fishing is unprofitable (Fig. 5).

**The effect of management context**

We ignore any potential costs of finer management such as management costs and transaction costs (Naidoo et al. 2006) because of the uncertainty and variability of such costs. Whether increasing cost with management scale leads to a maximum net fishery benefit (aggregate fishery profit minus costs following subdivisions) at an intermediate management scale will depend on the shape of the cost function (e.g., linear, exponential, or saturating relationship) compared to the autocorrelation-driven shape of the profit curve. For example, under the simplest possible case of a linear increase in cost with management scale, the maximum net fishery benefit is more likely to occur at an intermediate management scale for fisheries in an autocorrelated landscape (Fig. S4a; with its saturating profit function vs. scale relationship) than an uncorrelated landscape (Fig. S4b; with its near linear profit function vs. scale relationship, such that the resulting net benefit function will also be linear; see Appendix S3 for a detailed explanation).

The cost of coarser scale management in our model arises from the potential to assign fishing effort to unprofitable environmental patches, including non-habitat patches. However, note that the profit in any management unit is always nonnegative ( $\pi_{i,t} \geq 0$ ) under the optimal fishing strategies. Targeting of unprofitable locations may occur in reality due to ineffective initial searching by fishers for the harvested stock in a broader management region. Learning which regions are productive versus unproductive might be particularly slow, and therefore costly, if fishers do not share information (Allen, 2000) or population abundance varies in time.

In our model, this inefficiency is higher in the management with a coarser scale and it approaches to zero as the management scale becomes finer (Fig. 5). Explicitly incorporating the cost associated with the gap between ecological and management scales would require accounting for factors such as fisher behavior and institutional cooperation (Hilborn et al., 2005).

Ownership among management patches is also an important factor determining the relationship between fishery profit and management scale. White and Costello (2011) investigated the effect of size of the management unit for territorial user rights fisheries (TURF, i.e., management unit) with a two-patch model consisting of homogeneous environment. They concluded that optimal harvesting with sustainable, maximized yield occurs if all fish stay within TURF boundaries, such that TURF owners fully "own" all fish within their respective territories. This requires very small fish movement and/or very large TURF size, where any increase in fish mobility or decrease in TURF size that increases fish movement outside the TURF boundaries reduces ownership; if extreme, this reduced ownership can lead to overharvesting. Clearly, the difference the larger optimum management unit of White and Costello (2011) and our increasing profit with smaller management units is due to their incorporation of competition among fishermen, where assured ownership over the management unit determines the management success. In contrast, here we model the dynamics that would occur under federal level decision-making, cooperative management, or sole ownership, where competition among fishermen does not occur.

Some fishery management regimes have a hierarchical allocation of fishing effort where each decision-making sector has different management scales. For example, in Japanese and Chilean TURF system, federal level regulations in each management unit could further be arranged among local fishermen by a finer management unit (Makino and Matsuda 2005, Cancino et al. 2007). In fact, Hilborn et al. (2005) noted that the hierachical allocation of fishing effort caused serial depletion and collapse of California's abalone fisheries. Namely, state-level regulations create highly heterogeneous fishing pressure in the fishing ground, generating a mixture of depleted local populations and lightly exploited populations (Richards and Davis 1993, Karpov et al. 2000). The optimal spatial scale of management under such hierarchical management regimes, a management approach in between our analysis and White and Costello (2011), will ultimately depend on the combination of incentives at both the individual and federal levels. Note that hierarchical scales could be incorporated into our model, such as by applying zero fishing effort to a large cluster of non-habitat patches but allocating effort more finely in other patches in the FE strategy. If we allow a management strategy to apply heterogeneous management unit scales, the additional dimension to optimize over would likely lead to faster initial increases in, and earlier diminishing returns for, aggregate fishery profit as a function of the number of subdivisions.

Beyond reserves and harvested zones, small-scale spatial management has been increasingly relevant to marine systems (Shepherd 2003, Hilborn et al. 2005, Sanchirico and Wilen 2005). Marine population dynamics often occur on smaller spatial scales than the typical commercial fishery management scale of hundreds to thousands of kilometers inherent to regulatory institutions (Hilborn et al. 2005). A system of co-management, where the government and individual fishers both contribute to management decisions and implementation (Pomeroy and Williams. 1994, Pomeroy and Berkes 1997), would lead a finer management scale that better matches the biologically relevant scale and therefore can increase fishery profit, as suggested by our results. For example, in the co-managed Japanese TURF system, local communities submit individual management plans to the Prefecture for regional-level coordination (Hilborn et al. 2005).

Information availability can also affect optimal fisheries management and social decision-making (Andelman and Willig 2002, Richardson et al. 2006). We assumed implicitly perfect knowledge of managers in our model, as is often the case where managers intend to optimize the management outcomes based on metrics such as habitat quality (Neubert 2003, Sanchirico and Wilen 2005). Our model is most relevant to the situation where managers have a high degree of geographic and biological information. A high degree of information might be more common in systems of co-management, where fishermen typically collect finer-scale fisheries information because of the local-level management (Pomeroy and Berkes 1997), than the case of stock-level management modeled here. For example, in the case of fishing cooperative associations (FCA) in Japan (i.e., a co-management fishery), local, regional, and national governmental coordination in the design and implementation of fishery regulations (Lim et al. 1995) may allow a stock-level management institution to acquire fine resolution geographic and biological information. Technological innovations in marine spatial management can also facilitate collection of fine resolution data by a stock-level management institution. For example, remote sensing can map important ocean processes that influence species distributions, and geographic information systems (GIS) technology can help identify locations with essential habitat (Valavanis et al. 2004, 2008). More typically, given imperfect knowledge of demographic and biological parameters, our model might overestimate fishery profits, especially when managers apply a fine management scale, because of uncertainties in choosing appropriate patches. In this case, another trade-off, between flexible management and optimization errors, might arise.

**Conservation vs. fishery management goals**

Our analysis shows the significance of the choice of management scale in spatial fishery management: applying a different spatial scale of management alters the outcome for both in economic and ecological metrics, such as fishery profit, reserve fraction, fishing mortality rate, and population biomass. Specifically, a finer management scale allows fishermen to increase fishery profit effectively, but it causes a larger decline in biomass because of the more fine-tuned fishing (Figs. 3, 4), showing a tradeoff between economic values versus conservation value in spatial planning (also observed in White et al. 2012; Rassweiler et al. 2014). Hence, a careful consideration of both management goals and management scale is crucial for management success.

## Acknowledgements

This work was supported by Grant-in-Aid for Japan Society for the Promotion of Science (JSPS) Fellows to NT. We were also supported by Global Center of Excellence (GCOE) Program entitled "Asian Conservation Ecology as a basis of human-nature mutualism", JSPS. We thank N. Kumagai, L. W. Botsford, Y. Imai, H. Yokomizo, M. Springborn, J. N. Sanchirico, and Y Iwasa for their thoughtful comments.

# Supplementary Information

## Appendix S1: Optimization of the aggregate fishery profit

Here we describe the heuristic algorisms used to obtain local maximum fishery profit in each landscape for a given management scale.

*(i) Uniformed effort strategy*

The uniformed effort strategy (UE strategy) achieves an optimization of aggregate fishery profit by choosing $2^n+1$ parameters together so as to maximize the objective function at equilibrium $\Pi^*_{2^n} = \sum \pi^*_i$, where $2^n$ is the number of management patches and $i$ represents the environmental patch. First, $2^n$ binary parameters $(A_1, A_2, \cdots, A_{2^n})$ represent a reserve or fishing ground in each management patch, and the additional parameter $F$ represents the intensity of the fishing effort applied to all fishing grounds. For each value of intensity of fishing effort $F$, we calculate an optimal set of reserves and fishing grounds allocation $(A_1, A_2, \cdots, A_{2^n})$, and finally we select the set of $(A_1, A_2, \cdots, A_{2^n}, F)$ that shows the highest equilibrium aggregate fishery profit. We start with $F = 0.1$ and increase $F$ by an increment of 0.1 unless optimized fisheries profit goes to 0. We attain an optimal allocation of reserves and fishing grounds $(A_1, A_2, \cdots, A_{2^n})$ by applying the greedy algorithm, a procedure in which we start with each management patch designated randomly as either a reserve or fishing ground and then search across all of the management patches from the first management patch until we find a case in which switching a patch's designation increases the aggregate profit. After making this switch, we repeat the same procedure from the first management patch until we find an allocation of fishing grounds and reserves where no switch increases the aggregate fishery profit, following Rassweiler et al. (2012). We maintain $\Pi^*_{2^{n+1}} \geq \Pi^*_{2^n}$ by imposing multiple simulation runs if the condition is not satisfied. Comparison with other heuristic algorithms is discussed in more detail in Rassweiler et al. (2012).

*(ii) Fine-tuned effort strategy*

By natural extension of the conditional-multiple-start greedy algorithm discussed above, we attain an optimization of the aggregate fishery profit under fine-tuned effort strategy (FE strategy) by choosing $2^n$ parameters $(F_1, F_2, \cdots, F_{2^n})$, where $2^n$ is the number of management patches and $F_i$ represents fishing effort in management patch $i$. To find the optimal values of $\left(F_1, F_2, \cdots, F_{2^n}\right)$, we apply an analogue of the greedy algorithm and a local search algorithm with, if necessary, multiple trials with different initial conditions. We start with each management patch having a fishing effort level assigned from an independent and identically distributed random variable with range $F_i \in [0,1]$ with 0.1 step size, and we impose an amount and/or decrement of the intensity of fishing effort by an increment of 0.1 upon the first management patch and search across all of the management patches until we find a case in which an increment and/or decrement of the intensity of fishing increases the aggregate profit. Then we make the switch and repeat the same procedure from the first management patch until finding final allocation of fishing effort $\left(F_1, F_2, \cdots, F_{2^n}\right)$ where no switch improves the aggregate profit. In addition to the condition $\Pi^*_{2^{n+1}} \geq \Pi^*_{2^n}$, we maintain $\Pi^*_{2^n}$ of the FE strategy $\geq \Pi^*_{2^n}$ of the UE strategy by imposing multiple simulation runs if these conditions are not satisfied.

**Table S1** Red abalone (*Haliotis rufescens*) parameters. The same parameter values are used in the analysis in White et al. (2010).

| Parameter | Description | Value | Source* |
|---|---|---|---|
| $L_\infty$ | Maximum size | 19.24 cm | a |
| $k$ | Growth rate | 0.2174 cm/year | a |
| $a_0$ | Age at 0 cm | 0 year | a |
| $b_1$ | Coefficient in length-to-weight relationship | $1.69 \times 10^{-4}$ | b |
| $b_2$ | Exponent in length-to-weight relationship | 3.02 | b |
| $E_a$ | Fecundity-at-length | $15.32 L_a^{4.518}$ eggs | c |
| $a_{\max}$ | Maximum age | 30 years | d |
| $a_{\text{mat}}$ | Age at maturity | 3 years | e |
| $a_{\text{catch}}$ | Age available to fishing | 8 years (17.8 cm) | f |
| $M$ | Natural mortality rate | 0.15/year | g |
| $\bar{m}$ | Mean larvae dispersal distance | 5 km | f, h |

* Sources:

(a) Tegner, M. J., DeMartini, J. D. & Karpov, K. A. (1992). The California red abalone: a case study in complexity. In: *Abalone of the world: biology, fisheries, and culture*. (ed. Shepherd, S. A., Tegner, M. J. & Guzman del Proo, S. A.). Blackwell Science, Oxford, pp. 370–383.

(b) Ault, J. S. (1982). Aspects of laboratory reproduction and growth of the red abalone, *Haliotis rufescens Swainson* (Doctoral dissertation, Humboldt State University).

(c) Hobday, A. J. & Tegner, M. J. (2002). The warm and the cold: Influence of temperature and fishing on local population dynamics of red abalone. *Cal. Coop. Ocean. Fish.,* 43, 74-96.

(d) Leaf, R. T. (2005). Biology of the red abalone, *Haliotis rufescens*, in Northern

## Appendix S2: Sensitivity analysis

Here we show the results of a sensitivity analysis focused on the parameters with the greatest expected uncertainty (Fig. S1) and the results for the second species tested, namely Red Abalone (*Haliotis rufescens*; Figs. S2 and S3). The qualitative trends highlighted in the main text, such as the diminishing returns in aggregate fishery profit with finer management scale given habitat autocorrelation, are consistent across species and parameter values.

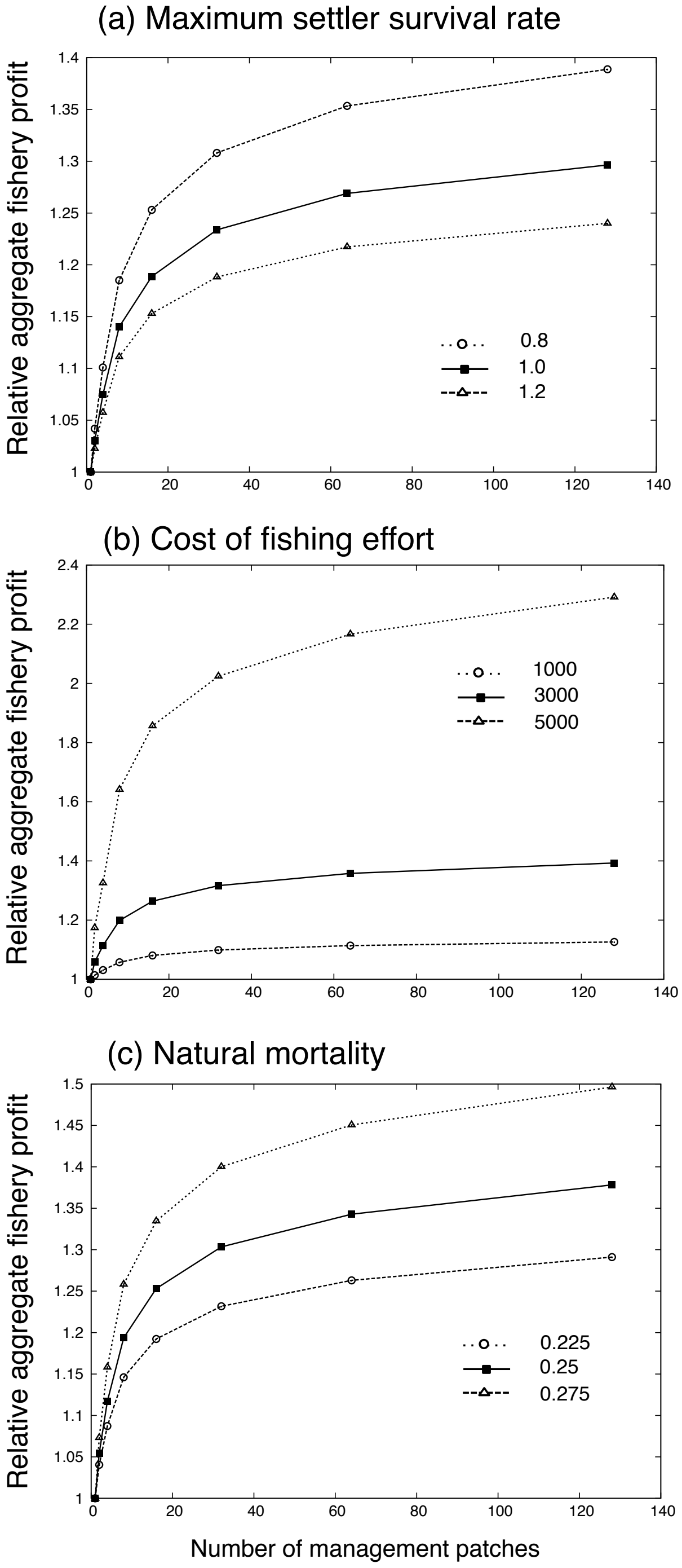

**Figure S1** Dependence of the relative aggregate fishery profit in an autocorrelated landscape ($\phi_1 = 0.9$) under the FE strategy, plotted relative to its value given one management patch, on (a) maximum settler survival rate $\alpha$, (b) cost of fishing effort $c$, and (c) natural mortality $M$. Red lines represents the value used in the main text.

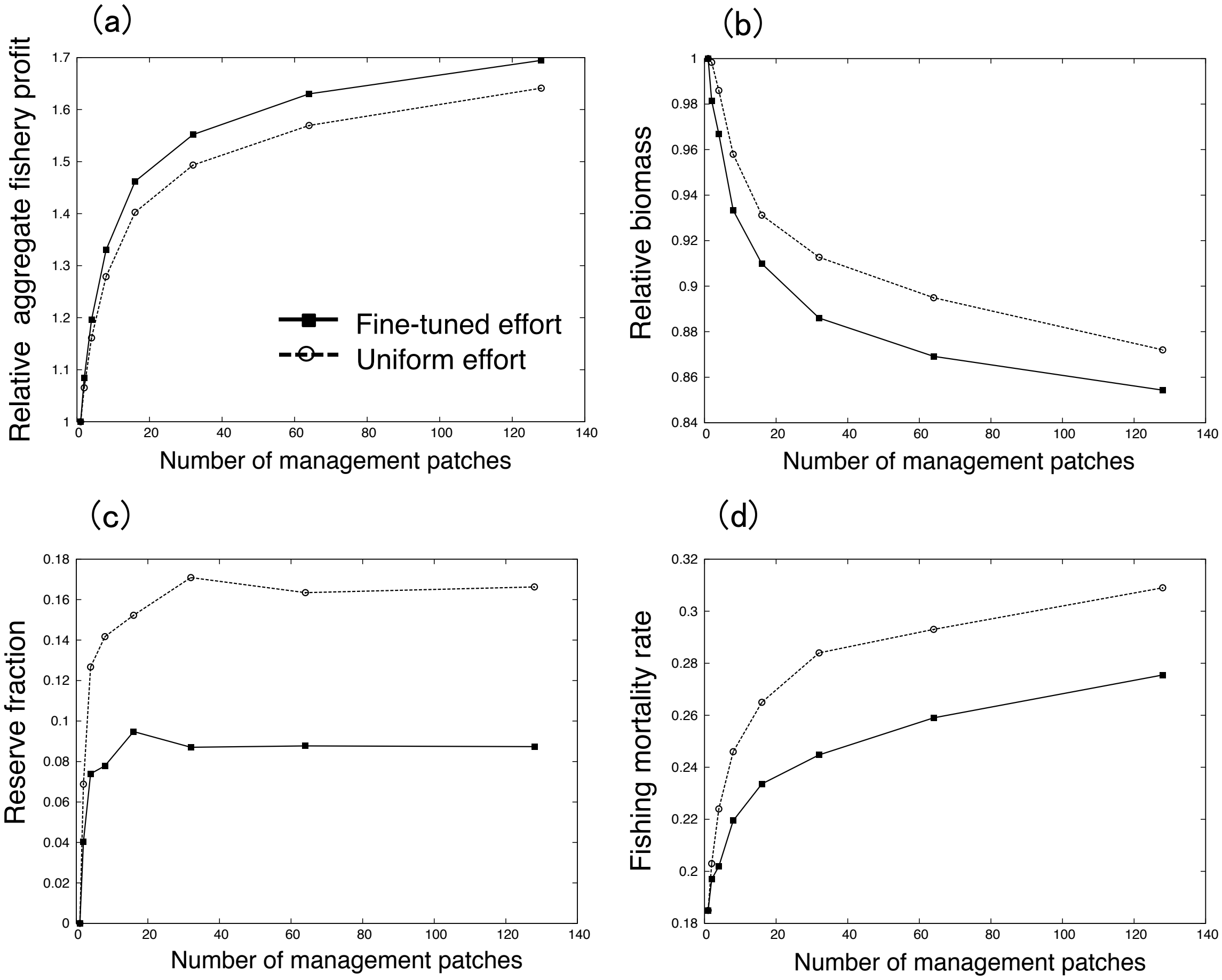

**Figure S2** The effect of management scale on red abalone (*Haliotis rufescens*) in an autocorrelated landscape under the FE strategy (squares, solid line) or the UE strategy (circles, dashed lines). Each panel shows (a) aggregate fishery profit, plotted relative to its value given one management patch, (b) biomass, plotted relative to its value given one management patch, (c) reserve fraction, and (d) average fishing mortality rate.

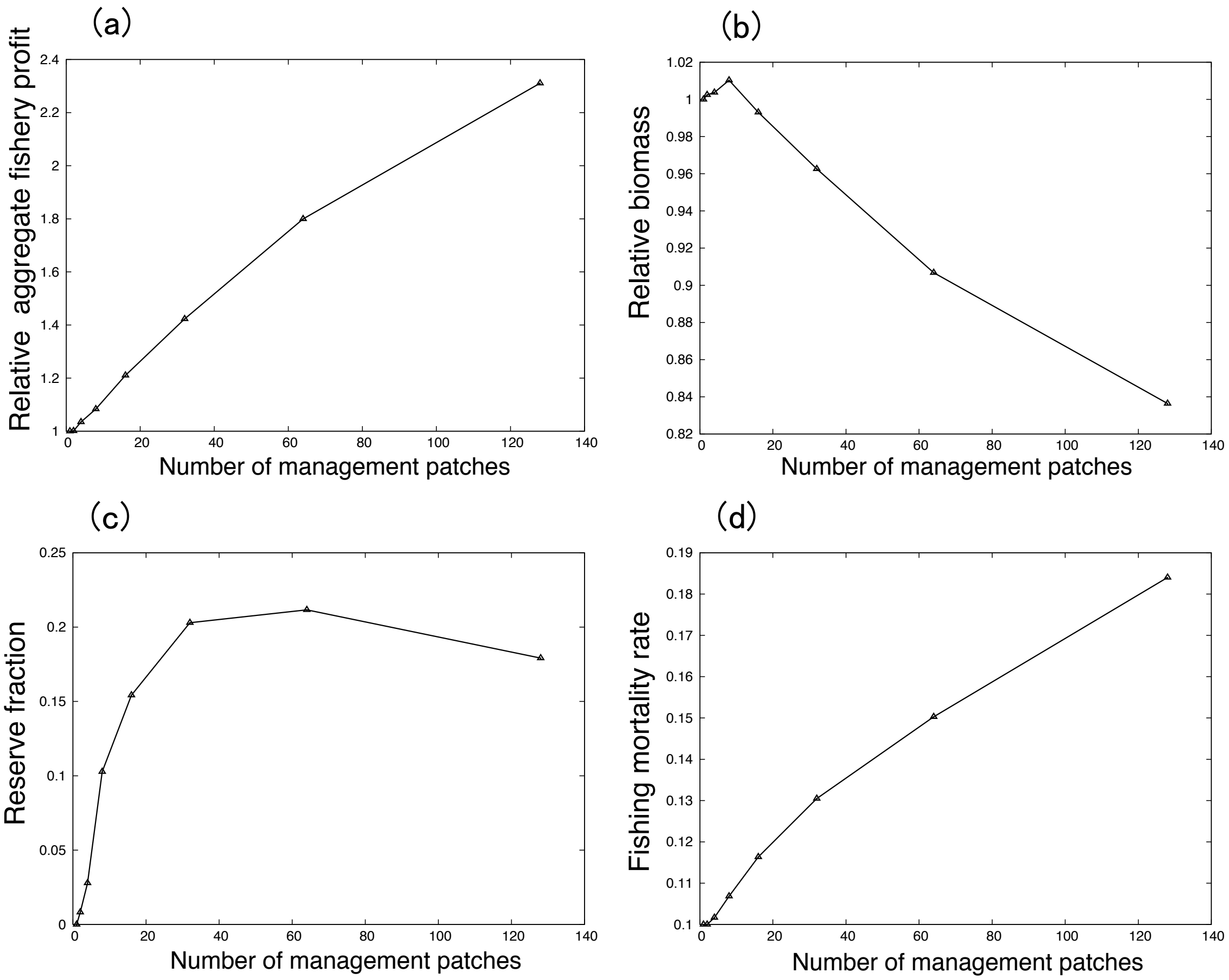

**Figure S3** The effect of management scale on red abalone (*Haliotis rufescens*) in an uncorrelated landscape under FE strategy. Each panel shows (a) aggregate fishery profit, plotted relative to its value given one management patch, (b) biomass, plotted relative to its value given one management patch, (c) reserve fraction, and (d) average fishing mortality rate.

## Appendix S3: Derivation of cost

As an example for how we might account for costs, let us assume that the cost of management increases with the number of management patches and is static over time. We may obtain a functional form $\alpha_1 N^{\alpha_2}$ as an additional cost of spatial fishery management, where $N$ is the number of management patches ($1 \le N \le 128$; see the main text) and both $\alpha_1$ and $\alpha_2$ are > 0. By subtracting this additional cost from Eq. 4 at equilibrium, we arrive at a net fishery benefit at equilibrium of $\hat{\Pi}^*_{2^n} = \sum \pi^*_i - \alpha_1 2^{n\alpha_2}$, where $2^n$ is a feasible number of management patches. Fig. S4 conceptually illustrates the net fishery benefit in the case where $\alpha_2 = 1$, i.e., assuming that the aggregated fishery profit is a liner function of the number of management patches $N$ in an uncorrelated landscape. In an autocorrelated landscape, if a cost of management $\alpha_1 2^{n\alpha_2}$ exceeds an aggregate fishery profit $\sum \pi^*_i$ in $0 < n \le 7$, the curve of the net fishery benefit $\hat{\Pi}^*_{2^n}$ is likely to have an intermediate optimum in the management scale that maximizes the net fishery benefit, $\hat{\Pi}^*_{2^n}$ (Fig. S4a). Alternatively, in an uncorrelated landscape, the net fishery benefit does not have an intermediate optimum (Fig. S4b). More generally (i.e., for all $\alpha_2 > 0$), in an uncorrelated landscape where the aggregated fishery profit is an approximately liner function of the number of management patches $N$, we can describe the aggregated profit with slope $A_1$ and intercept $A_2$: $A_1 N + A_2$. As defined above, the net fishery benefit is the aggregate fishery benefit minus the additional cost: $\hat{\Pi}^*_N = A_1 N + A_2 - \alpha_1 N^{\alpha_2}$. The second derivative of the net fishery benefit is then $d\hat{\Pi}^*_N / dN^2 = -\alpha_1 \alpha_2 (\alpha_2 - 1) N^{\alpha_2 - 2} > 0$ for $0 < \alpha_2 < 1$, and therefore is a convex function (Boccara 1990), implying that there is no intermediate optimum. In fact, the net fishery benefit can have a global maximum for $\alpha_2 > 1$ because of the concavity as the above condition suggests, but not for $N \ge 1$ where our analysis is focused. We can briefly explain this fact: In the domain $N \ge 1$ and for $\alpha_2 > 1$, the additional cost is the lowest at $N = 1$ and increases monotonically with $N$, and the net benefit function $\hat{\Pi}^*_N$, which is a linearly increasing function minus a monotonically and exponentially increasing function, is a monotonically decreasing function in $N \ge 1$, implying that it does not have an intermediate optimum. Therefore, an intermediate optimum in management scale given any cost to subdivision is unlikely to occur in an uncorrelated landscape but is likely to occur in an autocorrelated landscape.

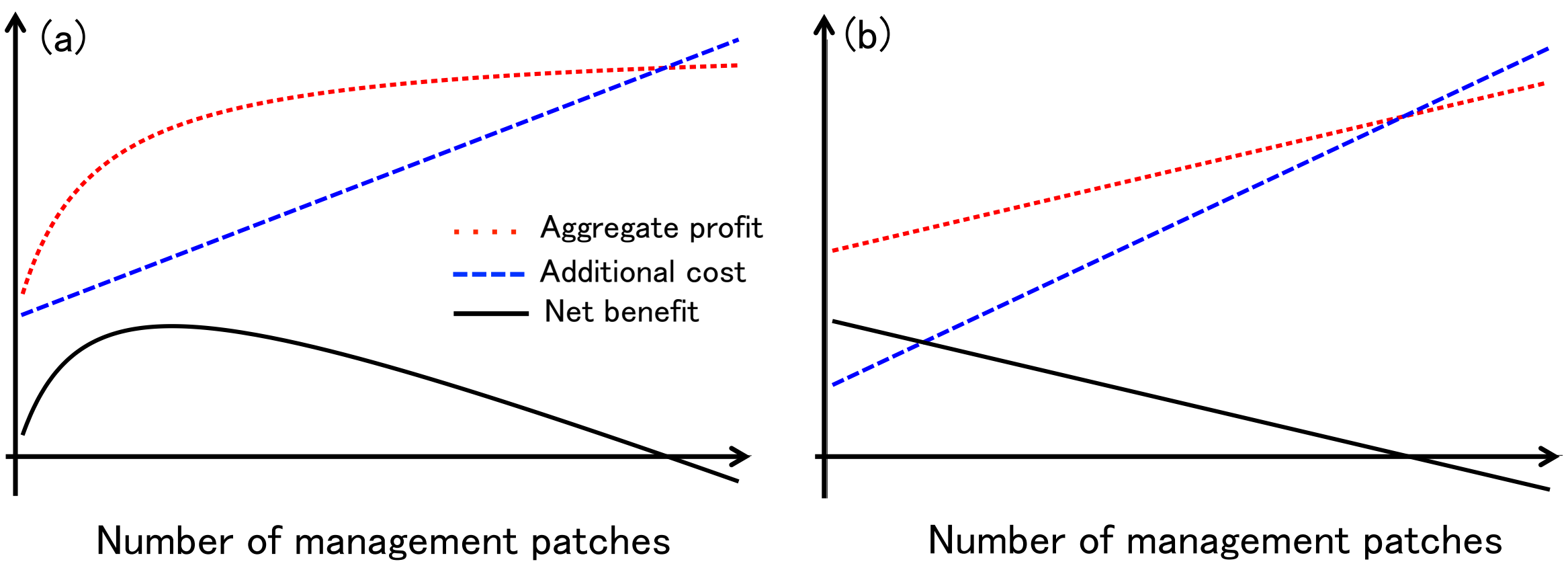

**Figure S4** Conceptual diagram of the net fishery benefit, which is the aggregate fishery profit (dashed red lines) minus an additional cost (dashed blue lines). For illustration, we assume that the cost of subdivision (dashed blue lines) increases linearly with an increasing number of management patches (e.g., as might occur due to increasing enforcement costs). (a) The net fishery benefit (solid black line) in an autocorrelated landscape has an intermediate optimum given the saturating profit function (dashed red line). (b) The net fishery benefit in an uncorrelated landscape does not have an intermediate optimum because the net benefit decreases monotonically with the number of management patches (solid black line).

## Appendix S4: Validity of the simulation settings

Here we verify that the number of trials used in our simulations is sufficient to capture qualitative trends. The qualitative relationship between aggregate fishery profit and the number of management patches is consistent regardless of the number of simulation trials used (25-150 trials), and the simulations converge in the range of the 100-trial results (Fig. S5a). The rapidly saturating relationship between reserve fraction and the number of management patches is consistent across realizations, while the variability within that relationship represents the stochastic effects of habitat patch selection (Fig. S5b; see Results: Management efficacy and parameter sensitivity).

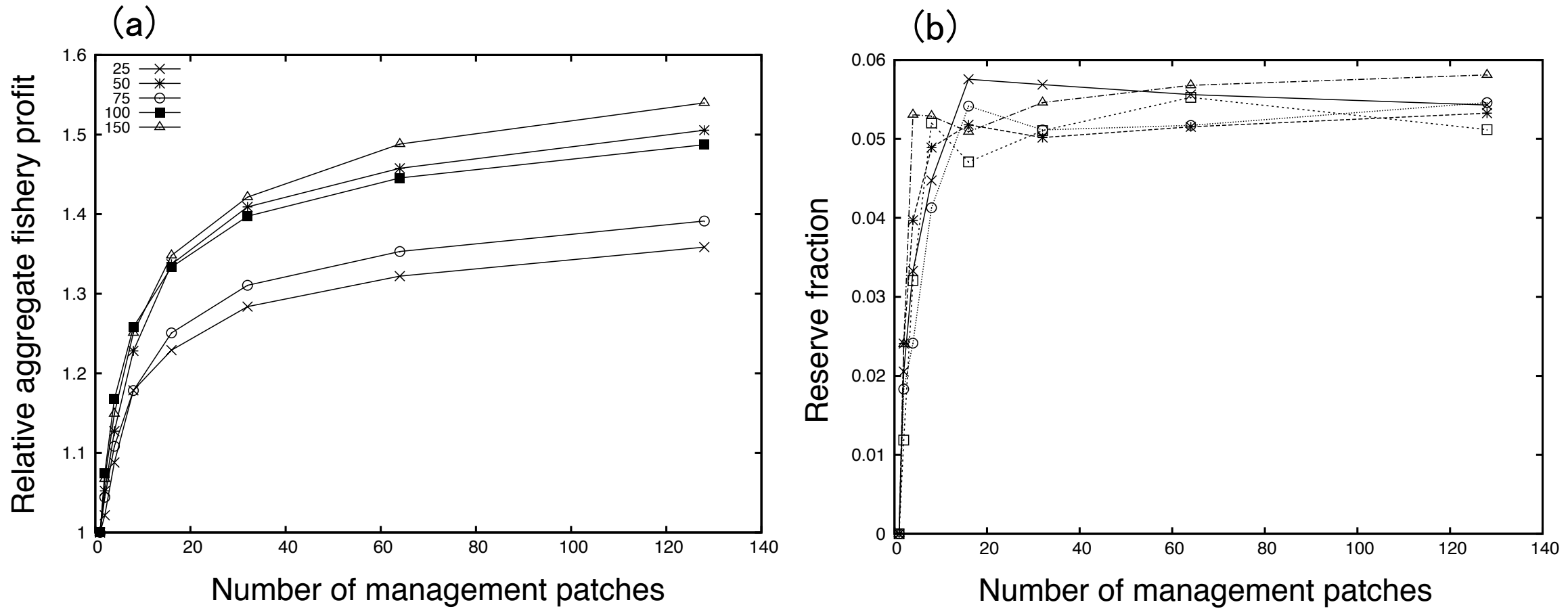

**Figure S5** Verification that 100 simulation trials captures qualitative trends for the fine-tuned effort strategy. (a) Aggregate fishery profit in an autocorrelated landscape ($\phi_1 = 0.9$) with various numbers of simulation trials. Each line, plotted relative to the value given one management patch, shows the average value of simulation trials: 25 (cross), 50 (star), 75 (circle), 100 (box), and 150 (triangle). (b) Realizations of reserve fractions of the five different hundred-time simulation trials.